\documentclass[%
 aip,
 amsmath,amssymb,
 reprint,%
]{revtex4-1}

\usepackage{graphicx}
\usepackage{dcolumn}
\usepackage{bm}

\usepackage[utf8]{inputenc}
\usepackage[T1]{fontenc}
\usepackage{mathptmx}
\usepackage{svg}
\usepackage{amsmath}
\usepackage{caption}
\usepackage{subcaption}
\usepackage{xspace}
\usepackage{adjustbox}

\usepackage{hyperref}
\hypersetup{colorlinks=true, linkcolor=blue, citecolor=blue, urlcolor=blue, unicode=false}

\begin{document}

\preprint{AIP/123-QED}

\title[]{A narrow bandwidth extreme ultra-violet light source for time- and angle-resolved photoemission spectroscopy}

\author{Qinda Guo}

\author{Maciej Dendzik}
\author{Antonija  Grubi\v{s}i\'{c}-\v{C}abo}
\author{Magnus H. Berntsen}
\author{Cong Li}
\author{Wanyu Chen}
\affiliation{ 
Department of Applied Physics, KTH Royal Institute of Technology, Hannes Alfv\'{e}ns v\"{a}g 12, 114 19 Stockholm, Sweden
}%
\author{Bharti Matta}
\author{Ulrich Starke}
\affiliation{Max Planck Institute for Solid State Research,
Heisenbergstra{\ss}e 1, 70569 Stuttgart, Germany}
\author{Bj$\mathrm{\ddot{o}}$rn Hessmo}
\author{Jonas Weissenrieder}
\author{Oscar Tjernberg}%
 \email{oscar@kth.se}
\affiliation{ 
Department of Applied Physics, KTH Royal Institute of Technology, Hannes Alfv\'{e}ns v\"{a}g 12, 114 19 Stockholm, Sweden
}%

%

\date{\today}

\begin{abstract}
Here we present a high repetition rate, narrow band-width, extreme ultraviolet photon source for time- and angle-resolved photoemission spectroscopy. The narrow band width pulses $\Delta E=9, 14, 18$ meV for photon energies $h\nu=10.8, 18.1, 25.3$ eV are generated through high harmonic generation using ultra-violet drive pulses with relatively long pulse lengths (461~fs). The high harmonic generation setup employs an annuluar drive beam in a tight focusing geometry at a repetition rate of 250 kHz.  Photon energy selection is provided by a series of selectable multilayer bandpass mirrors and thin film filters, thus avoiding any time broadening introduced by single grating monochromators. A two stage optical-parametric amplifier provides $< 100$~fs tunable pump pulses from 0.65~$\mu$m to 9~$\mu$m. The narrow bandwidth performance of the light source is demonstrated through angle-resolved photoemission measurements on a series of quantum materials, including the high-temperature superconductor Bi-2212, WSe$_2$ and graphene.
\end{abstract}

\maketitle

\section{\label{sec:Introduction}Introduction\protect }
Angle-resolved photoemission spectroscopy (ARPES) is one of the prime experimental techniques for determining the electronic band structure in crystalline solids. Apart from being able to capture the bare energy-momentum dispersion of electronic states in a material, ARPES directly probes the single-particle spectral function which contains information regarding band dispersion and many-body interactions\cite{damascelli2003angle}. For this reason, ARPES has become an indispensable tool for understanding emergent phenomena in quantum materials (QM) where quasiparticles or collective excitations are core ingredients and naturally invoke a description beyond the simple non-interacting single-particle picture \cite{hengsberger1999electron,lashell1996spin}. 

\begin{figure}[htp]
\includegraphics[scale=0.1]{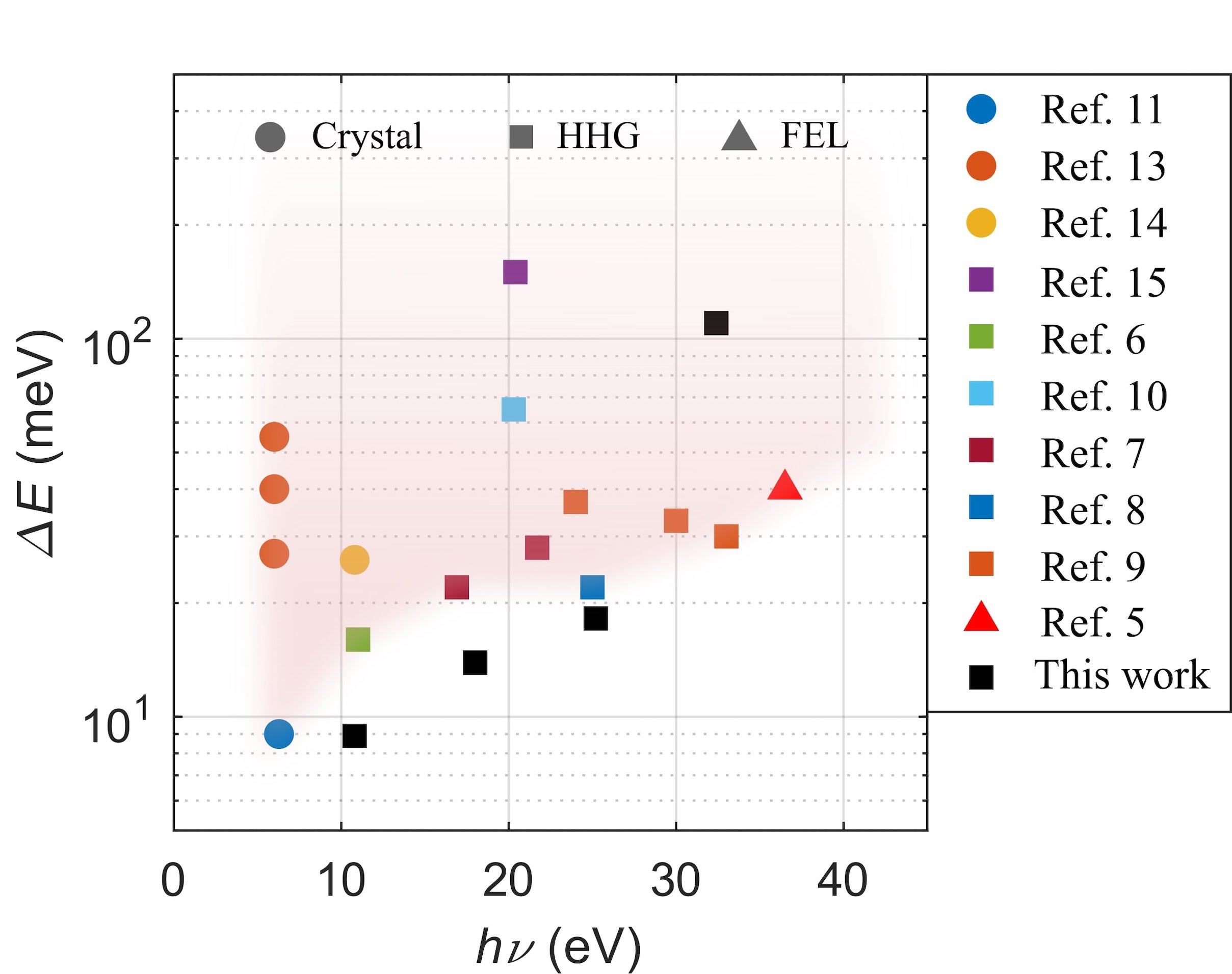}
\caption{\label{fig:survey}State-of-the-art energy resolutions of current time-resolved ARPES setups plotted as a function of the photon energy in semi-logarithmic scale. The different techniques for generating the probe source are indicated by the shape of the markers.}
\end{figure}

Over the past decades, there have been great improvements in synchrotron-based light sources which, together with new and more versatile electron analyzers, have pushed the energy and momentum-space resolutions in ARPES experiments to unprecedented levels \cite{HighResARPES_Review_Iwasawa2020,kutnyakhov2020time}. In parallel to this development, ARPES has taken a leap into the time-domain due to the advent of femtosecond high-power lasers that have enabled ultrafast extreme-ultraviolet (XUV) light sources based on high-harmonic generation (HHG) in noble gases\cite{lee2020high,cucini2020coherent,mills2019cavity,sie2019time,corder2018ultrafast} or non-linear crystals\cite{parham2017ultrafast,kuroda2017ultrafast,gauthier2020tuning,peli2020time}. While several successful implementations of time- and angle-resolved photoemission spectroscopy (tr-ARPES) systems have been demonstrated, the technique continues to rapidly evolve and there is still much progress to be made in terms of increased repetition rate, photon flux, improved time and energy resolutions, photon energy and momentum range coverage, as well as pump versatility. To date, most HHG based tr-ARPES systems\cite{petersen2011clocking,eich2014time,rohde2016time,buss2019setup,puppin2019time} have focused on reaching a high time resolution with limited repetition rates since this requires relatively modest average laser powers and therefore is more accessible Some state-of-the-art examples of tr-ARPES setups are shown in Fig. \ref{fig:survey}. 

Ideally, for conducting tr-ARPES experiments one would like an ultrafast high-repetition rate laser-based light source that simultaneously provides high temporal and energy resolutions through short pulses with a narrow bandwidth, respectively, as well as providing sufficiently energetic photons to cover a large region of momentum space. A high repetition rate of the probe, on one hand, mitigates space charge induced broadening\cite{SolidStateDynamics_book,zhou2005space,passlack2006space} of the ARPES spectra by reducing the number of photoelectrons generated per pulse while maintaining high average count rates. On the other hand, an increased repetition rate reduces the maximum achievable pulse energy which affects the laser’s ability to efficiently drive the HHG process. Furthermore, shorter pulses provide higher peak power for the same pulse energy and lead to both improved time resolution and HHG efficiency, but at the expense of reduced energy resolution. In addition, one would ideally also prefer a tunable pump, covering a large wavelength range, so that the energy can be selected to optimize excitation efficiently of the process of interest.

Clearly, there is an inherent trade-off between the achievable time and energy resolution in tr-ARPES, intimately linked to the fact that short pulses in the time-domain have a broad spectral content. Designing a light source for tr-ARPES, therefore, requires careful considerations in order to reach the combination of repetition rate, available photon energies, and time and energy resolutions that match the time and energy scales of the electron dynamics that one ultimately wants to study.

Here, we present a laser-based XUV source that is integrated into the BALTAZAR facility\cite{berntsen2011experimental} for tr-ARPES. The light source has been designed to achieve high energy resolution at high repetition rates in the ARPES setup in order to resolve detailed electronic structure features in quantum materials such as superconducting gaps in superconductors. The key parameters of the light source and photoemission setup are listed in Tab. \ref{tab:parameters}. The light source uses a tabletop laser (Amplitude, Tangor 100), delivering 461~fs long pulses at a wavelength of 1030~nm, which are then frequency tripled, through second harmonic and sum frequency generation in non-linear crystals, to 343~nm. The 343~nm light pulses are in turn used to produce vacuum-ultraviolet (VUV) photons with energies 10.8~eV, 18.1~eV, 25.3~eV and 32.5~eV at a repetition rate of 250~kHz through high harmonic generation in argon gas. Static ARPES measurements of the Fermi edge in polycrystalline Au using 10.8~eV, 18.1~eV and 25.3 eV photon energies yield a total experimental energy resolution of 9~meV, 14~meV and 18~meV, respectively. This energy resolution is the total system resolution, including contributions from the light source, analyzer, space charge and stray fields. As such, it puts an upper bound on the spectral width of the harmonics. Time-resolved ARPES measurements on graphene using probe energies of 18.1~eV and 25.3~eV exhibit a time resolution of 204~fs and 165~fs, respectively. The photon energies available through the HHG process give a momentum coverage that extends beyond the first Brillouin zone in most QM and the high detection efficiency of the angle resolved time-of-flight (ARTOF) analyzer allows for measuring the electron dynamics over a two-dimensional region in momentum space in parallel, without the need for rotating the sample or deflecting the emitted photoelectrons. The 250~kHz repetition rate permits space charge broadening to be kept below a detectable level, while maintaining high average count rates on the detector. The HHG drive laser optically seeds a second 280~fs 1030~nm laser (Amplitude, Tangerine HP2) that drives a two-stage optical parametric amplifier (OPA; Fastlite, TwinStarzz) which in turn provides pump pulses in the $0.65-9$~$\mu m$ wavelength range. Together, the wide combination of pump and probe wavelengths permits tailoring of the pump-probe combination for the particular system under study.

\begin{table}
\caption{\label{tab:parameters} Key parameters for the tr-ARPES setup.}
\begin{ruledtabular}
\begin{tabular}{lcccr}
\qquad \qquad \qquad \qquad \quad System performance\\
\hline
Photon energy (eV) &10.8 &18.1&25.3&32.5\\
Energy resolution (meV)\footnotemark &9 & 14 & 18 & 111\\
Time resolution (fs) & - & 204 & 165& - \\
XUV pulse duration (fs) & - & 178 & 131 & - \\
Time-bandwidth product (meV$\cdot$fs) & - & 2492 & 2358 & - \\
Photon flux on sample (photons/s)\footnotemark & $2e11$&$8e9$&$7e8$&$7e7$\\
XUV spotsize($\mu m^2$)                                        & \multicolumn{4}{c}{96 (H) $\times$ 85 (V)}\\
Repetition rate (kHz)\footnotemark & \multicolumn{4}{c}{250}\\
\end{tabular}
\footnotetext[1]{Total system energy resolution deduced from ARPES measurements.}
\footnotetext[2]{Determined based on the drain current measurement from tantalum foil, with a picoampmeter (Keithley, 6485). Yield efficiencies of 0.08, 0.20, 0.10 and 0.08 for the $3^{\mathrm{rd}}$, $5^{\mathrm{th}}$, $7^{\mathrm{th}}$  and $9^{\mathrm{th}}$ harmonics, respectively, are estimated from Refs. \onlinecite{feuerbacher1972experimental,diaz2019experimental}.}
\footnotetext[3]{For time-resolved measurements. Repetition rate for static ARPES is practically tunable from 250~kHz to 1~MHz.}
\end{ruledtabular}
\end{table}

\section{System overview}
\begin{figure*}[htp]
\includegraphics[scale=0.8]{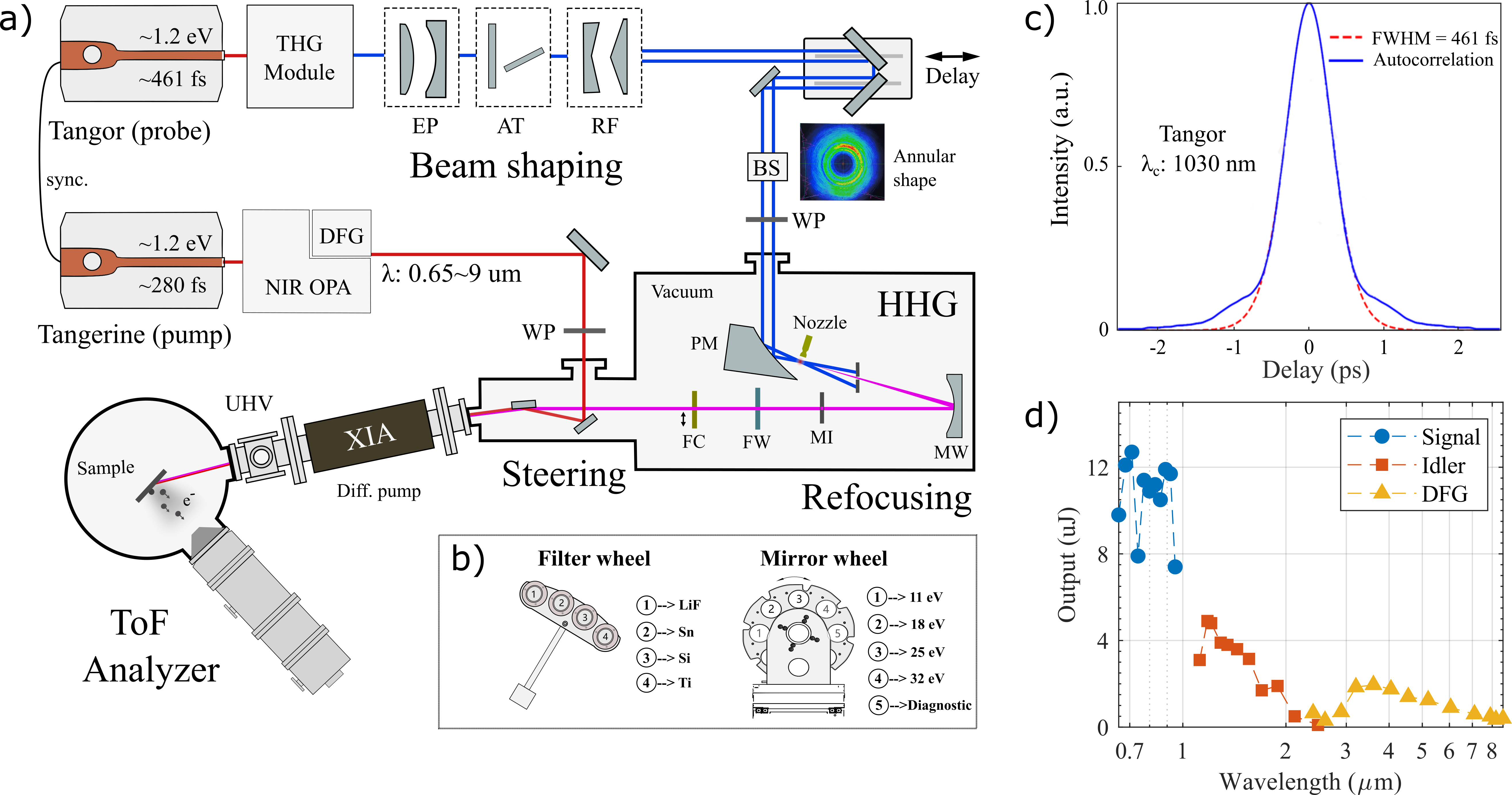}
\caption{\label{fig:Setup}Schematic overview of the HHG-based time-resolved ARPES setup. a) The layout of the setup, showing the pump and probe lines. Abbreviations: EP - Expander, AT - Attenuator, RF - Reflective axicons, BS - Beam stabilization, WP - Waveplate, PM - Parabolic mirror, MW - Mirror wheel, MI - Motorized iris, FW - Filter wheel, FC - Flux check. b) Drawings of the mirror- and filter-wheels that are used for wavelength selection. c) Temporal profile of the $\sim$461-fs-long ultraviolet (UV) pulse used for driving the HHG probe-line. d) OPA performance, depicting the average pulse energy as a function of the output wavelength in semi-logarithmic scale.}
\end{figure*}

Figure \ref{fig:Setup}a) presents a schematic overview of the tr-ARPES setup. In this section, following the beam path, we describe the principle segments of the system.

\subsection{The probe-line}
A high-power femtosecond laser (Amplitude, Tangor 100) is used as the source, providing infrared (IR) pulses centered at 1030~nm, with an adjustable repetition rate from single shot up to 40~MHz. The maximum average-output-power exceeds 100~W. At 250~kHz repetition rate, the energy per pulse is 300~$\mu J$. We stress the $\sim$ 461~fs long pulse length adopted here, as shown in Fig.\ref{fig:Setup}b), as the key point for achieving high energy resolution. Following the laser amplifier, there is a third harmonic generation (THG) module which uses nonlinear crystals to convert the IR into 343 nm (3.6 eV). The efficiency for this process is $\sim 30 \% $, and it corresponds to a maximum output power of $\sim$30~W and the pulse energy of 88~$\mu J$. For a practical pump-probe scheme, the repetition rate has to be the result of a compromise. Considering measurement statistics and resolution, one favors higher repetition rate, as space charge can be mitigated and count rates kept high. However, excited-state relaxation time and thermal diffusion of pump energy in the sample after excitation set an upper bound on the repetition rate - as does the available laser power and resulting photon flux. The effect of thermal load and photo yield is also sample dependent which makes optimizing these parameters a non-trivial problem. In the present case, 250~kHz was chosen as a reasonable trade off for a large range of samples and pump conditions. The choice of OPA repetition rate was made at the design stage and the working point of 250 kHz has therefore not been subject to experimental optimization.

A critical step in the present setup is beam shaping before the high harmonics generation. The aim is to transform the intensity distribution of the beam from a Gaussian profile to an annular shape, in which the intensity is near zero in the central region. This approach permits the drive beam to be separated from the generated harmonics along the beam propagation direction. This forms the basis for the use of refocusing mirrors and filters to select the photon energy without having to handle the full power of the drive beam. Specifically, the 3.6~eV (343 nm) Gaussian drive beam with diameter $\sim$3~mm (1/$e^2$), is first expanded, using a convex and a concave dielectric mirror, to a diameter of $\sim$6 mm. This pre-expansion is done in order to reduce the power load on the downstream optical elements, thus reducing thermal wavefront distortion and damage risk. The former is a particular problem for transmission optics such as wave-plates and windows. A power attenuator (EKSMA) is placed after the beam expander, and is used to modulate the pulse energy externally, without altering the running conditions of the drive laser, which can result in changes in pulse characteristics and beam pointing. Following the attenuator, a pair of convex and concave reflective axicons (Mflens Natsume) are used to transform the Gaussian beam into an annular beam. The inner and outer diameters of the annular beam are 3~mm and 9~mm, respectively. The beam profile after the reflective axicons is shown in the inset of Fig.\ref{fig:Setup}a). Note that there is remaining intensity in the beam center, and this intensity is dumped by reflecting the beam off a mirror with a center hole at a 45$^\circ$ angle prior to passing the beam into the HHG chamber.

A delay stage (Newport, ESP301) carrying two plane dielectric mirrors, is used to generate the optical delay between the probe and the pump. The delay stage is placed in the probe-line, since the output of OPA covers a wide wavelength range from visible to mid-infrared, which have different beam paths, thus making it more challenging to implement the delay stage in the pump path. In order to maintain HHG, probe- and pump-beam stability, both beam paths are actively stabilized using piezo mirrors and beam sensors (Thorlabs PDA90A). 

\subsection{High harmonics generation}
HHG is by now a well established technique for upconverting visible or infrared laser light into the vacuum-ultraviolet (VUV) and X-ray regions\cite{LaserHHG_KepteynMurnanePRL1996, HHG_CoherentSoftXrays_Murnane_PRL1997, HHG_CoherentXrays_WaterWindow_Krausz_Science1997}. The physical process behind HHG is well understood and well described theoretically\cite{gorlach2020quantum}. For a HHG source to be used as a photon source for ARPES it should ideally be bright, have a high repetition rate (> 100 kHz), have a narrow line width and be tunable. Since HHG generation requires drive laser intensities on the order of $I_L \approx 10^{14} W/cm^2$, the drive laser pulse length is usually kept short in order to limit the necessary pulse energy. In the case of high repetition rates on the order of kHz, this also limits the necessary average power. In the present case, the goal has been to achieve a high repetition rate photon source that can serve as a narrow bandwidth source with a line width on the order of 10 meV without the need for monochromatization. In order to achieve this, several trade-offs had to be made. The drive pulse length has to be substantially longer than commonly applied, and the pulse energy has to be kept low in order for the average power to remain reasonable. Long pulses of low pulse energy need a very tight focus to achieve the necessary intensity given above, and as a result the gas target pressure needs to be high\cite{rothhardt2014absorption}. Even under these conditions, the efficiency of the HHG process drops dramatically for longer pulses, being an order of magnitude lower for 450 fs pulses as compared to 45 fs pulses\cite{shiner2009wavelength,hadrich2016single}. To address this issue, a cascaded approach is used in which the drive laser is upconverted to the 3rd harmonic, and this harmonic is then used to drive the HHG process. The cascaded approach significantly improves the HHG efficiency since the HHG generation efficiency scales as $\lambda^{-(5-6)}$\cite{tate2007scaling,comby2019cascaded}. The use of a shorter drive wavelength, however, reduces the available harmonics as seen from the single atom high energy cut-off relation\cite{HHG_UP_theoryCalc_Krause_PRL1992} $h\nu_{cutoff}= I_p+3.17U_p$, where $h$ is Planck's constant, $I_p$ the ionization potential and $U_p$ the quiver energy of the electron. The cut-off energy scales as $\lambda^2$ since the quiver energy of the electron $U_p \propto I_L \lambda_L^2$, with $I_L$ being the drive laser intensity and $\lambda_L$ the wavelength, leading to a dramatically reduced cut-off energy for shorter wavelengths\cite{HHG_Ne_Ar_LHullier_PRA1993}.

\begin{figure}[htp]
\includegraphics[scale=0.75]{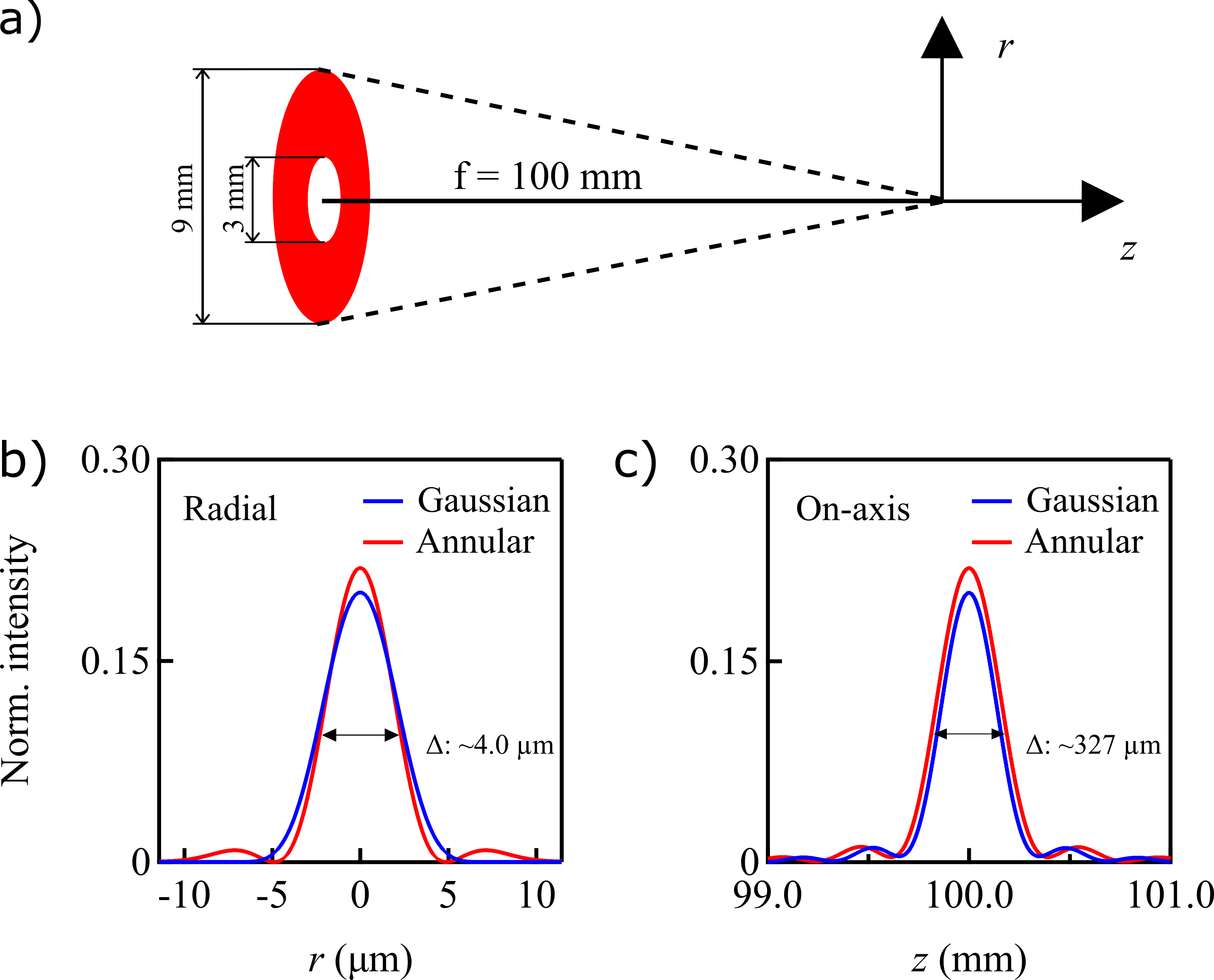}
\caption{\label{fig:HHG}Numerical simulation of the focusing annular beam. a) The geometry of the annular beam with the dimensions. b) The radial and c) on-axis intensities across the focus for a  9~mm (1/$e^2$) diameter Gaussian beam and an annular beam with 3~mm inner diameter. The focal length is set to 100~mm.}
\end{figure}

In order to be able to spatially separate most of the drive beam from the generated HHG beam, the intensity profile of the drive laser beam is transformed into an annular shape using reflective axicons as described above. The annular beam is focused into an argon gas jet\cite{HHG_GasJetOptimisation_KatsumiPRA2000} using an off-axis parabolic mirror (Thorlabs, MPD169) resulting in a Gaussian annular beam focus of $\sim$6.8 $\mu$m diameter (1/$e^2$) and $\sim$164 $\mu$m Rayleigh length. The theoretical radial intensity profile at the focus as well as the on-axis intensity profile across the beam focus are shown in Fig.\ref{fig:HHG}, demonstrating that the annular intensity profile will not significantly change the quality of the focus compared to a Gaussian beam. 

When focusing the beam in the gas target, in order to achieve phase matching at the focus while minimizing re-absorption, a very high density gas target is required with a size that extends on the order of the Rayleigh length. In the present setup, a high density gas jet is used as the target which is provided by a 150 $\mu$m diameter de Laval nozzle with a 50 $\mu$m throat diameter. The experimentally determined phase matching pressure is achieved with a 4.5 bar injection pressure. In order to maintain the best possible background pressure, the injection nozzle faces a counter nozzle with a 2 mm diameter opening situated approximately 200 $\mu$m in front of the injection nozzle. The counter nozzle is pumped by a high capacity scroll pump (Edwards, XDS35i) and the vacuum chamber itself is pumped by a 500 l/s turbo pump (Pfeiffer, HiPace 700). The resulting vacuum chamber background pressure is $\sim 7\times10^{-7}$ mbar and $\sim 6\times10^{-4}$ mbar during HHG generation. 

\subsection{Monochromator}
Gratings are commonly adopted in ARPES setups to select the desired wavelength. In the present setup where the available wavelengths are well separated by the HHG process we instead do energy selection by employing band-pass mirrors and thin-film filters. This configuration brings advantages in at least three aspects. Firstly, we can extract the photon flux of the direct reflection, instead of the limited flux of the 1st diffraction order from a grating. Secondly, the stretching of the pulse from diffraction can be avoided, which is important for a time-resolved setup. Thirdly, it helps with the XUV imaging properties as there is no angular dispersion introduced.

For compactness and ease of alignment, a near normal incidence geometry with multiple mirrors mounted on a revolving wheel was chosen, c.f. Fig.\ref{fig:Setup}b). The rotatable wheel is placed after the HHG interaction region and has a set of mirrors optimized for the different wavelengths. All mirrors are spherical in order to refocus the diverging HHG beam onto the sample position. The wheel is on a high-precision translation stage (Smaract), which enables a nanometer-scale fine tuning of its position. One of the mirrors is SiC coated, providing high reflectivity for the 5th harmonic (18.1 eV). Two multi-layer coated mirrors are used for the 7th (25.3 eV) and 9th (32.5 eV) harmonic and one $MgF_2/Al$ mirror for the 3rd harmonic (10.8 eV). The mirrors for the 7th and 9th harmonics are coated by Ultrafast Innovations and provide peak reflectivities of $41\%$  and $27\%$, respectively. The angle of incidence of the refocusing mirrors is $\approx$1$^\circ$ and the radius of curvature is 1000~mm. 

A set of thin-film filters (Lebow) is used to clean the spectra from residual intensity remaining from other harmonics than the one selected as well as from the drive beam. This assembly is illustrated in Fig.\ref{fig:Setup}b). Matching the mirror selection the filter wheel contains the following options: LiF (10.8 eV), Sn (18.1 eV), Si (25.3 eV) and Ti (32.5 eV) thin films for selecting the four different harmonics, as well as an option of using an Al filter (cuts harmonics below 15 eV as well as the drive beam). The thickness of the filters, with the exception of LiF, is less than 200 nm, which improves transmission but makes them sensitive to high power loads. A motorized iris (Standa) is placed before the filters to further reduce the power load on the filters. Additionally, it can be used to regulate the photon flux without changing the power of the drive laser, thus keeping the HHG generation conditions fixed, which allows space charge effects to be directly monitored.

The selected harmonic is guided into the ARPES analysis chamber by a steering mirror. This mirror is mounted in a 19$^\circ$ grazing incidence angle and is coated with gold to provide high reflectivity in the XUV, as well as MIR, wavelength range. The drain current from the steering mirror can be monitored during experiments and used as a photon flux reference. The steering mirror is used to steer the beam onto the sample, as well as align the beam spot on sample with the focus of the electron energy analyzer. 

\subsection{The pump-line}
An additional femtosecond laser (Amplitude, Tangerine HP2) is used as the pump light source. It is an Ytterbium-Doped Fiber Amplifier (YDFA) laser and delivers the same central wavelength of 1030 nm as the Tangor laser used for the probe line. The two lasers are optically synchronised by sharing a common oscillator. The pulse picker of the Tangerine is triggered by a pulse generator (Quantum composer, 9200) to ensure that the Tangerine (pump) constantly picks the same pulse as the one picked by Tangor (probe) from the pulse train that the oscillator yields. The Tangerine output wavelength is 1030~nm (1.2 eV), with the pulse length of $\sim$280 fs and a maximum output power that exceeds 50 W. The Tangerine is used to drive the OPA which in turn delivers 3 tunable output modes: signal (0.65~$\mu$m - 0.95~$\mu$m), idler (1.15~$\mu$m - 2.4~$\mu$m) and difference frequency generation (DFG; 2.5~$\mu$m - 9~$\mu$m). The performance of the OPA is presented in Fig.\ref{fig:Setup}d), where the adjustable wavelength range and the pulse energy for each mode is given. As noted above, the choice of a 250~kHz repetition rate for the OPA was made in the design phase and is practically non-tunable. Note that due to the strong absorption of mid-infrared in air, the beam path of the pump line is fully enclosed and purged with nitrogen gas. The pump beam is coupled into the probe beam path by a 45$^{\circ}$ mirror with a center hole through which the probe beam can pass. This makes the pump and probe beams propagate close to colinear through the last section of the beam path leading to the sample, and allows for simultaneous adjustment of the position of both beams on the sample using the aforementioned steering mirror. When focusing the pump onto the sample in the analysis chamber of the ARPES setup, the spot size is $\sim$0.8 mm for the signal, and $\approx$ 2mm for the idler. The DFG beam size has not been characterized due to the difficulty of observing wavelengths in the IR range, but is expected to be 3-4~mm in the present focusing geometry. The pump-pulse duration is compressed to below 100~fs.

\subsection{Photoemission setup}
The photoemission setup consists of the analysis chamber, a preparation chamber, where sample preparation and characterisation can be done, and a load lock for fast sample entry. The details of the entire chamber layout and the functionalities can be found in Ref$\cite{berntsen2011experimental}$. Briefly, the functions of the preparation chamber include sputtering and annealing option for the sample cleaning, low-energy electron diffraction (LEED) for structure and quality determination of the sample surface, and thin-film deposition by electron-beam evaporation with up to three source cells.

The analysis chamber is capable of maintaining a base pressure of < 1 $\times$10$^{-10}$ mbar. The analysis chamber is connected to the photon beamline via a differential ion pump (XIA, DP-03). This provides a windowless line-of-sight transition from high vacuum (HV) in the photon beam line to ultra high vacuum (UHV) in the ARPES analysis chamber. The windowless solution has the advantage that it provides rapid switching between photon energies. A solution that uses a series of window valves, where different thin filters act as the vacuum barrier between the chambers, was initially considered but was deemed less flexible and robust. The current solution would, for example, permit future filter-less operation if other means for harmonic selection is developed, or additional harmonics are added. A motorized 4-axis manipulator (SPECS) is mounted on the analysis chamber and equipped with a closed-cycle cryostat (ARS, 4K). The lowest sample temperature that can be reached is $\sim$8 K. 

The ARTOF analyzer (SPECS, Themis 1000) is a line of sight analyzer consisting of an electrostatic lens system, and a Delay-Line-Detector (DLD, Surface Concept, DLD4040). The lens system can provide several imaging modes such as direct imaging or angle resolving modes for ARPES measurements. The DLD is synchronized to the photon pulse by a fast photodiode. The flight time and position data of the photoemitted electrons acquired by the DLD is converted into a three-dimensional data set of $k_x$, $k_y$ and $E_k$, where $k_x,k_y$ are crystal momentum coordinates and $E_k$ the kinetic energy of the photoelectron. This three dimensional, in parallel, data collection is the major advantage of the ARTOF as compared to a hemispherical analyzer and allows for an efficient acquisition of ARPES spectra over an extended area in momentum space without the need for rotating the sample or deflecting the photoelectrons.

\section{\label{sec:Performance}Performance\protect }
In this section, we show results from incorporating the HHG light source into the ARPES setup. We characterize two important parameters of the combined system, namely the XUV spot size at the sample position in the photoemission analysis chamber and the achievable total energy resolution in ARPES measurements using the different harmonics of the light source. We then present a few examples of ARPES measurements on quantum materials, thus directly demonstrating the practical capabilities of the source.

\subsection{XUV spot size}

\begin{figure}[htp]
\includegraphics[scale=0.45]{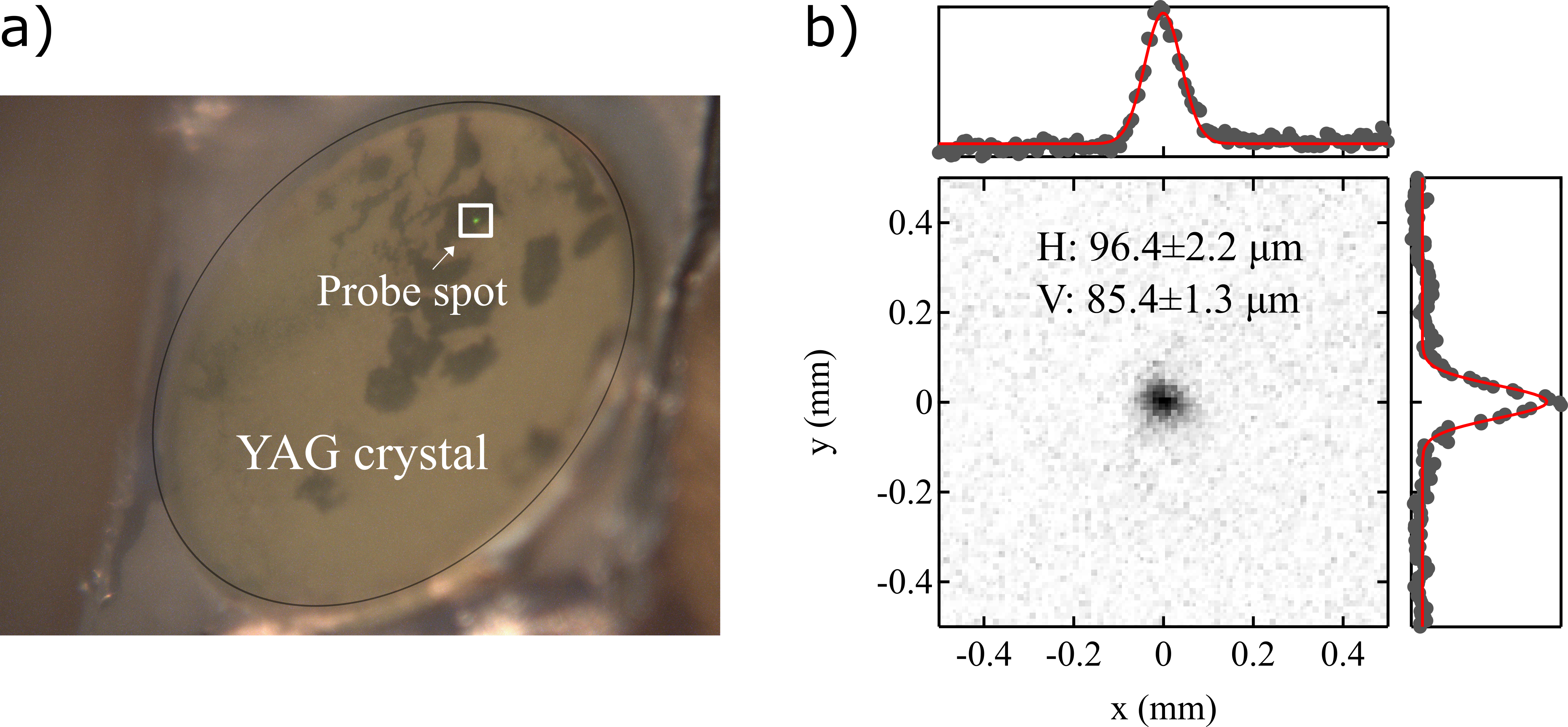}
\caption{\label{fig:SpotSize}Spot size measurement for the 7th harmonic (25.3 eV). a) Camera image of the probe beam on a YAG crystal which is placed at the sample analysis position in the ARPES chamber. The spatial scale of the image in the horizontal and vertical directions are calibrated using the motorized sample-manipulator displacement. b)  Gaussian fits of the vertical and horizontal profiles of the spot, yielding a spot size of approximately 96~$\mu$m and 85~$\mu$m in the horizontal and vertical directions, respectively.}
\end{figure}

The spot size of the 7th harmonic (25.3 eV) was determined to be $\sim$ 96 $\mu m$ $\times$ 85 $\mu m$, from photoluminescence on a YAG (Y$_3$Al$_5$O$_{12}$) crystal. The YAG crystal was placed at the sample analysis position in the ARPES chamber. The scale of the camera pixels in both directions was calibrated to the manipulator displacements, which can be precisely controlled using stepper motors. Figure \ref{fig:SpotSize}a) shows the YAG crystal together with the photoluminescence of the 25.3 eV probe spot. Figure \ref{fig:SpotSize}b) shows a zoom-in of the beam spot and the results of fitting the beam intensity distribution to a Gaussian profile along the horizontal (H) and vertical (V) directions, respectively. The beam size differs between the horizontal and vertical directions as expected from the off-normal incidence geometry but has overall a profile very close to that of a two-dimensional Gaussian profile. 

\subsection{Experimental energy resolutions}

\begin{figure}[htp]
\includegraphics[scale=0.7]{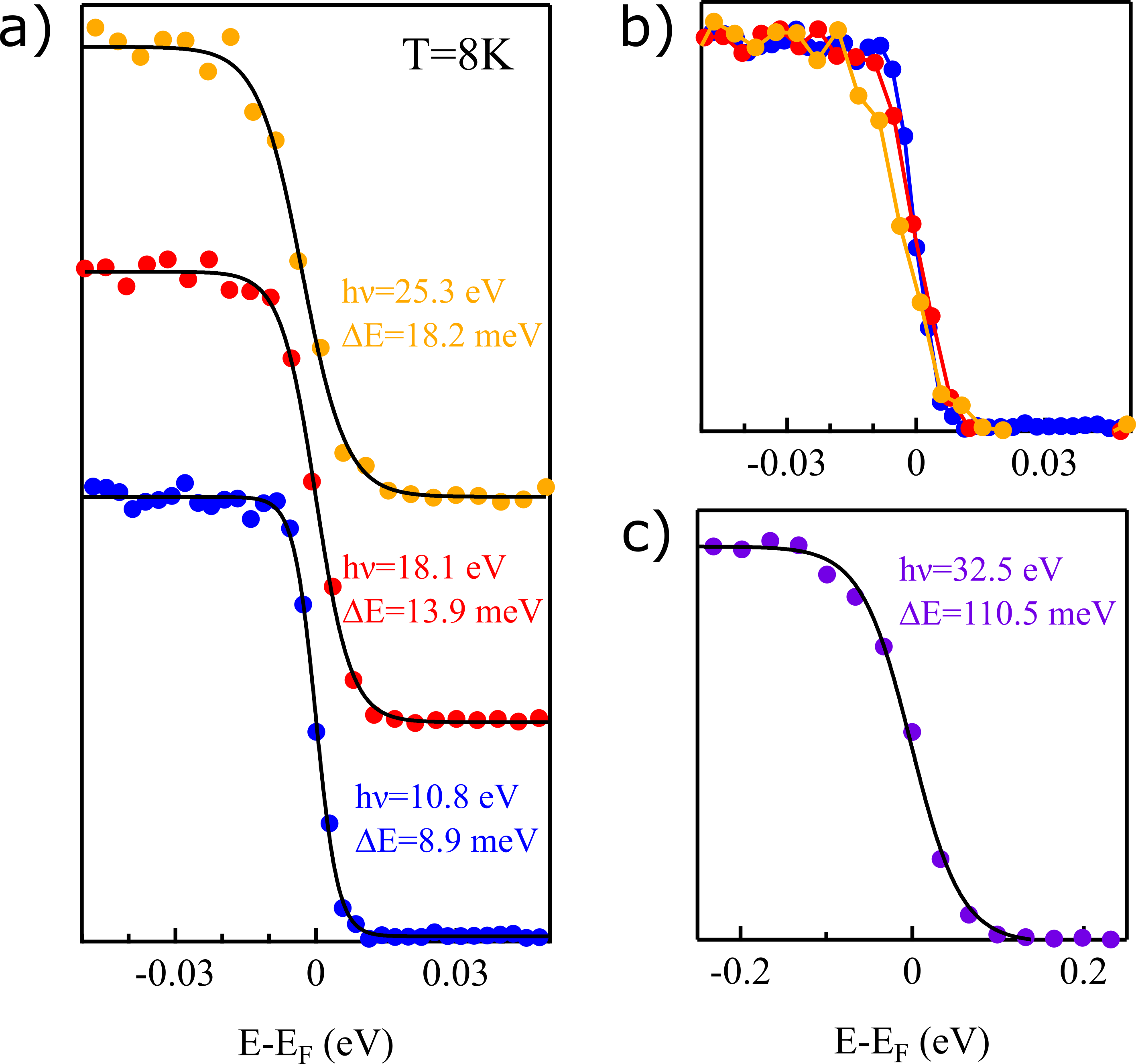}
\caption{\label{fig:Au_FL}Fermi edge measurements on polycrystalline Au taken at 8 K. a) Data acquired at 10.8~eV (blue), 18.1~eV (red) and 25.3~eV (yellow) plotted with a vertical offset. Solid black lines are fits using a convolution of the temperature dependent Fermi-Dirac function with a Gaussian function, where the full-width at half maximum of the Gaussian represents the overall system energy resolution. b) The same data without vertical offset. c) Results for 32.5 eV. Purple dots represent data, black line is the fit.}
\end{figure}

The experimental energy resolution of the system is characterized by Fermi-edge measurements on polycrystalline gold. Fresh gold is evaporated in the preparation chamber onto a gold foil mounted on a copper sample holder and transferred \textit{in situ} into the analysis chamber. This way, the achievable energy resolution for available harmonics was determined. Figure \ref{fig:Au_FL} shows the raw data (dots) from the measurements at the different harmonics together with Fermi-edge fits. The fitted curve consists of a convolution of the temperature-dependent Fermi-Dirac distribution and a Gaussian-profile, the latter representing the system energy resolution. The temperature used for the measurements and the fitting is 8~K. Overall, the 3rd harmonic (10.8~eV) shows the ability of reaching an energy resolution of 8.9 meV. The 5th (18.1~eV) and 7th (25.3~eV) harmonic yields energy resolution of 13.9~meV and 18.5~meV, respectively. This overall energy resolution contains contributions not only from the harmonic line width but also analyzer resolution, space-charge effects and stray fields. The majority contribution to the energy broadening is believed to come from the linewidth of the harmonics in view of the fact that the analyzer has previously demonstrated resolution better than 5~meV  \cite{berntsen2011experimental}, and that we do not observe any improvement of resolution or shift of Fermi edge if the photon flux is decreased. This is further supported by simulations for the expected analyzer resolution which show an expected resolution of 0.9~meV, 0.9~meV, and 6.8~meV for the 3rd, 5th and 7th harmonic Fermi-edge measurement, respectively. We note that the intrinsic linewidth of the driving laser is 3.9~meV, which corresponds to 0.37~nm at 343.4~nm wavelength. Therefore, the overall energy resolution of our system is mainly limited by the HHG process, during which the pulse experiences temporal compression and energy broadening. Included in Fig.\ref{fig:Au_FL} is also a Fermi edge measurement using the 9th harmonic (32.5~eV), which exhibits an energy resolution of 110.5~meV. This harmonic has very low intensity and the fitted Gaussian width in this case is not limited by the bandwidth of the harmonic but rather the analyzer resolution ($\sim$68.4~meV) as well as space charge broadening due to the presence of lower harmonics in the beam during the measurement. The very limited photon flux of this harmonic also prevents time-resolved measurements from being performed.

\subsection{ARPES test cases}
\subsubsection{Resolving the superconducting gap}

\begin{figure*}[htbp]
\includegraphics[scale=0.67]{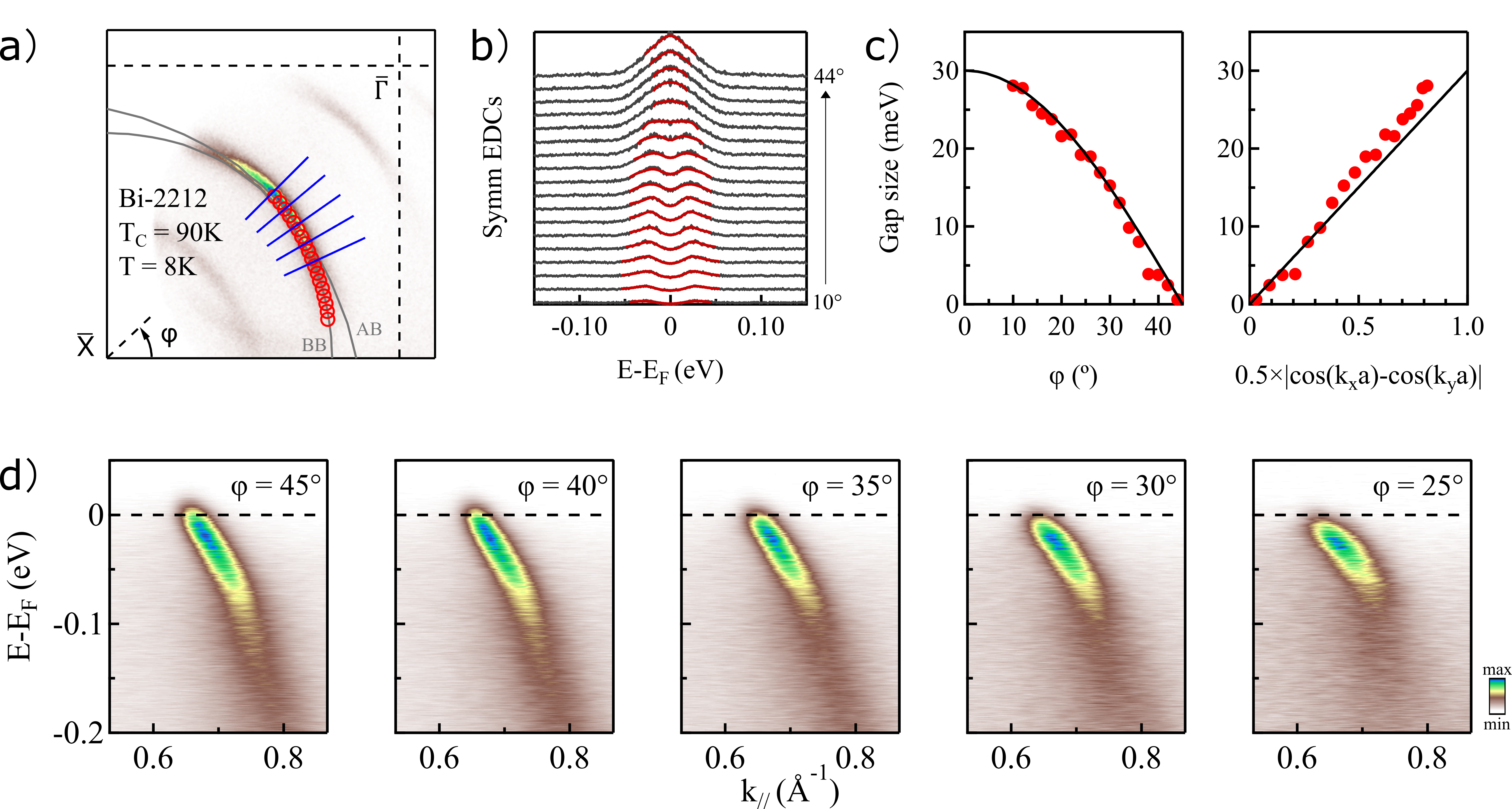}
\caption{\label{fig:BISCO}Low temperature (8~K) static ARPES measurement of the high-$T_\mathrm{c}$ cuprate superconductor Bi-2212 using 18.1 eV photons. a) Constant energy contour at 20~meV below the Fermi level. b) Symmetrised EDCs (black curves) extracted at the position of the red circles in a) along with fits to the EDCs (red lines). c) The superconducting gap as a function of the FS angle and 0.5$\times$$|\mathrm{cos}(k_\mathrm{x}a)-\mathrm{cos}(k_\mathrm{y}a)|$. d) Example slices taken at various FS angles, as indicated by the blue lines in a).}
\end{figure*}

The first test case consists of static measurements on the copper-based high-$T_\mathrm{c}$ superconductor Bi$_2$Sr$_2$CaCu$_2$O$_{8+x}$ (Bi-2212), to show the capability and feasibility of resolving the band structure in the whole first Brilloune zone, as well as the superconducting gap, with a HHG source. Since the discovery of high-$T_\mathrm{c}$ superconductors (HTS) decades ago\cite{bednorz1986possible}, this group of materials has attracted considerable attention. Among the HTS, the copper-based compounds (cuprates) are typically known for their high transition temperature and the comparative simplicity of their layered crystal structure. A complete understanding of the mechanism(s) underlying superconductivity in these systems is still lacking despite immense experimental and theoretical efforts\cite{imada1998metal,damascelli2003angle,sobota2021angle}. Although HHG-based light sources have progressed rapidly during recent years, access to the superconducting gap has been limited for HHG-based ARPES due to the relatively large bandwidth of these setups as compared to synchrotron radiation based setups. However, HHG is so far the most practical approach for ultrafast XUV source generations, which is favorable for time-resolved ARPES, especially considering the achievable time resolution and availability of lab-based systems. Figure \ref{fig:BISCO} shows static ARPES results for an optimally doped Bi-2212, measured with a photon energy of 18.1~eV. The data were acquired at a sample temperature of 8~K, well below the superconducting transition temperature ($T_\mathrm{c}=~$90~K). 

The wide-angle mode (WAM) of the ARTOF analyzer, with an acceptance angle of $\pm15^\circ$, was used to collect the data. Figure \ref{fig:BISCO}a) shows a constant energy contour at 20~meV below the Fermi level. The complete nodal to anti-nodal section is within the analyzer acceptance window. The splitting of the main band at the off-nodal direction, into a bonding (BB) and an anti-bonding (AB) band, can be resolved, and the red circles in Fig.\ref{fig:BISCO}a) are fit of the bonding band. Apart from the main band, the super-structure and shadow bands are also observed with a relatively lower intensity. Figure \ref{fig:BISCO}b) presents symmetrized energy distribution curves (EDCs) taken along the bonding band for Fermi surface angles ($\phi$) in the range 10$^\circ$ to 44$^\circ$ at $k$-space positions given by the fitted red circles in Fig.\ref{fig:BISCO}a). The red lines in Fig.\ref{fig:BISCO}b) are fits of the symmetrized EDCs using a Norman function convoluted with a Gaussian\cite{norman1998phenomenology}. The superconducting gap is determined as the difference in peak positions of the fitted phenomenological form. The gap size is plotted as a function of Fermi surface (FS) angle and 0.5$\times|\mathrm{cos}(k_\mathrm{x}a)-\mathrm{cos}(k_\mathrm{y}a)|$ in Fig.\ref{fig:BISCO}c). The black line represents the $d$-wave form, $30\times\mathrm{cos}(2\phi)$. The determined gap size and shape of the optimally doped Bi-2212, agrees well with previously published result\cite{sun2018temperature}. The near FS dispersion for a few selected FS angles, as indicated by blue lines in Fig.\ref{fig:BISCO}a), are presented in Fig.\ref{fig:BISCO}d), where the phonon-coupling induced kink ($E_\mathrm{b}\sim$70~meV) and the evolution of mode coupling strength in momentum can be clearly resolved. The presented data undeniably demonstrates the capability of the present setup to resolve and study the momentum dependence of the near-FS electronic structures and superconducting gaps in cuprate systems with future applications for the study of ultrafast dynamics in high-$T_\mathrm{c}$ superconductors\cite{zonno2021time}. 

\subsubsection{Momentum coverage}

\begin{figure*}[]
\includegraphics[width=.85\textwidth]{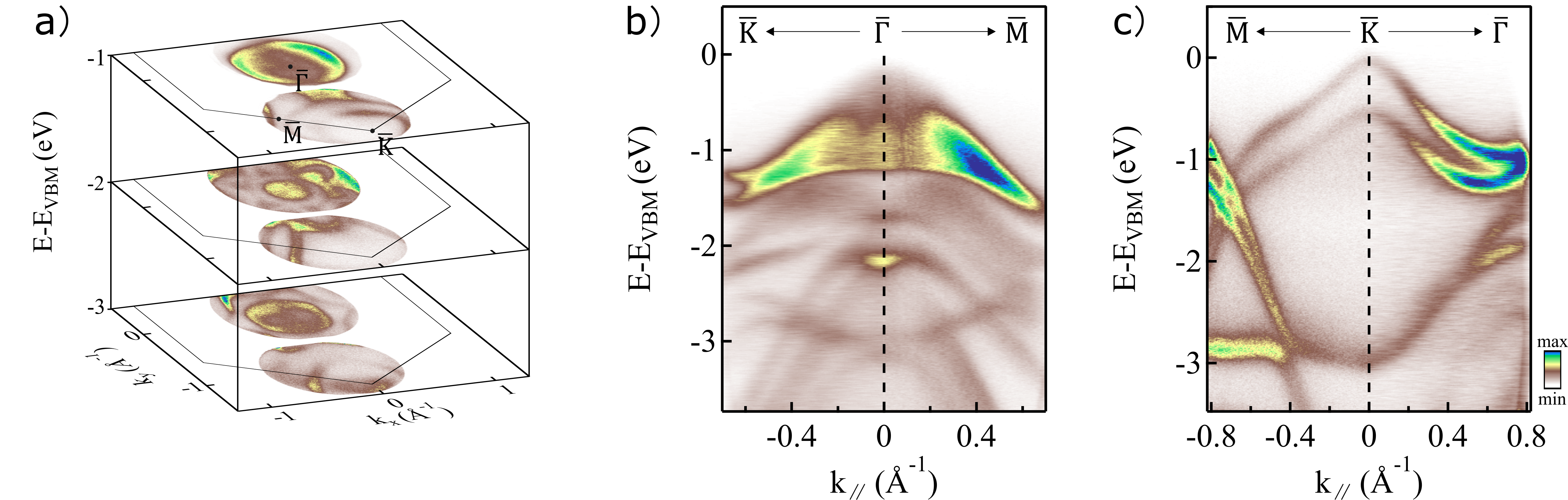}
\caption{\label{fig:WSe2_Bulk}Static WSe$_2$ band structure measured at room temperature with 25.3~eV photon energy. a) Constant energy contours taken at the $\bar{\Gamma}$ and $\bar{K}$ points and at 1, 2 and 3 eV below the valence band maximum. WSe$_2$ band structure b) at the $\bar{\Gamma}$ point along the $\bar{K}-\bar{\Gamma}-\bar{M}$ direction and c) at the $\bar{K}$ point along the  $\bar{M}-\bar{K}-\bar{\Gamma}$ direction. A valence band spin-orbit splitting of 514~meV at the $\bar{K}$ point is clearly visible in c). The binding energy scale is set to 0~eV at the valence band maximum at $\bar{\Gamma}$.}
\end{figure*}

To showcase measurements at large in-plane momenta, we use 2H-WSe$_2$ as an example, a member of the transition metal dichalcogenide (TMDC) family. 2H-WSe$_2$ (from now on referred to as WSe$_2$) is a semiconducting TMDC with an indirect gap of 1.25 eV \cite{2H_WEs2_IndirectGap, WSe2_BandGap} that retains bulk inversion symmetry while still exhibiting a large spin polarization of its bulk electronic states\cite{2H_WSe2_PDCKing}. Due to the presence of large spin polarisation in inequivalent $K$ and $K'$ valleys \cite{WSe2_BerryCurvature_ParkPRL2018, 2H_WSe2_PDCKing}, WSe$_2$ is a prospective candidate for spin- and valleytronic devices\cite{Valleytronics_rev}, making it scientifically and technologically an interesting system. The WSe$_2$ sample was cleaved \textit{in situ} using the top-post method and measured at a pressure of $1\times10^{-10}$~mbar. The analyzer WAM mode was used also in this case, together with an off-normal emission geometry to reach the $\bar{K}$ point of the first Brillouin zone of WSe$_2$ (1.3 {\AA}$^{-1}$). Measurements were done using a photon energy of 25.3~eV, which allows us to access the $\bar{K}$ point within our available polar rotation range and has favourable matrix elements for WSe$_2$ at both $\bar{\Gamma}$ and $\bar{K}$ points. The multi-layer refocusing mirror (coated for 25.3~eV) and an Al filter were employed to clean the spectra of higher and lower harmonics. Measurements were performed at room temperature, and total recording time was 6 hours. Figure \ref{fig:WSe2_Bulk}a) shows the constant energy contour at the $\bar{\Gamma}$ and $\bar{K}$ points at 1, 2 and 3 eV below the valence band maximum (VBM), while Fig.\ref{fig:WSe2_Bulk}b) and c) show example cuts through the data volume around the $\bar{\Gamma}$ and $\bar{K}$ points along the high-symmetry directions $\bar{K}-\bar{\Gamma}-\bar{M}$ and $\bar{M}-\bar{K}-\bar{\Gamma}$, respectively. The VBM for bulk WSe$_2$ is located at the $\bar{\Gamma}$ point, and the energy axis is referenced to the VBM. Valleys of WSe$_2$ are located at the $\bar{K}$ point, and a spin-orbit splitting of 514~meV at the $\bar{K}$ point is clearly resolved. These results show the feasibility of acquiring high statistics data at large in-plane momenta. The high detection efficiency and momentum coverage of the ARTOF analyzer becomes apparent at higher order harmonics where a large fraction of the Brillouin zone can be covered in one single measurement\cite{TOF_vs_Hemisphere_Mac_RevSciInstr2020}.

\subsubsection{Time-resolved experiment}

\begin{figure*}[htp]
\includegraphics[scale=0.86]{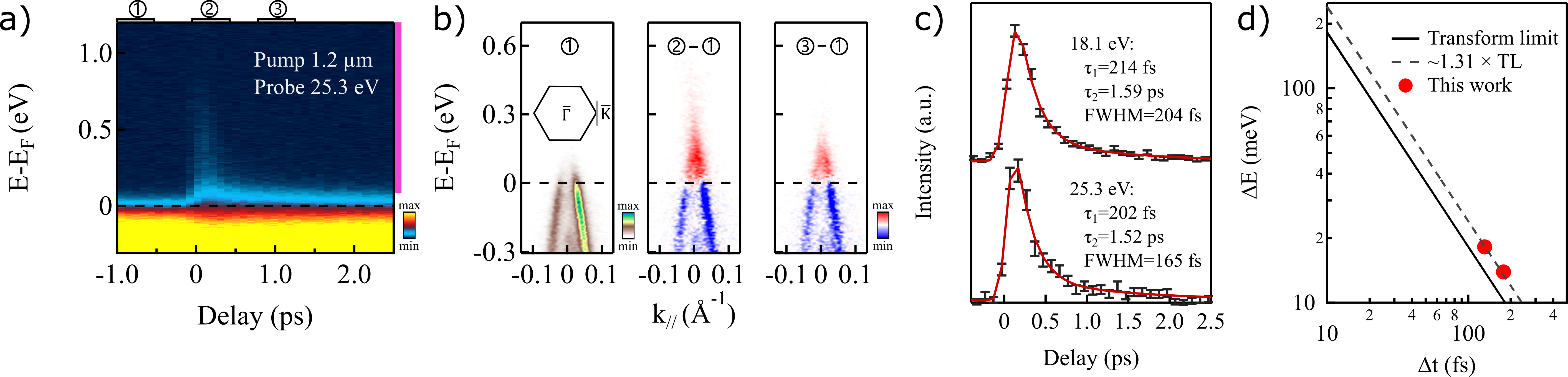}
\caption{\label{fig:TR_GR}Pump-probe measurements on p-doped graphene measured with 25.3~eV photons. a) Ultrafast dynamics of the graphene, excited with a pump beam of 1.2 $\mu$m wavelength and $\sim$100~fs pulse duration. Purple vertical line indicates the integration range in energy for the corresponding delay curve displayed in panel c). b) From left to right: energy dispersion cuts from a static measurement without pump (\textcircled{1}), at the excitation time ($t_0$, \textcircled{2}-\textcircled{1}) and 1~ps after $t_0$ (\textcircled{3}-\textcircled{1}). c) Fit to the decay data. The decay data is taken from an integration of the energy window marked by the purple line on the right side of a). The fitting profile is a 2-component exponential decay curve convoluted with a Gaussian function. d) Energy resolution ($\Delta E$) versus time resolution ($\Delta t$) for the photon energies of 18.1~eV and 25.3~eV, the solid line shows the theoretically resolution limit assuming a Fourier transform limited Gaussian pulse.}
\end{figure*}

In order to determine the temporal resolution reachable with the current source, p-doped graphene was used as a test sample. P-doped graphene has fast enough intrinsic dynamics to reflect the system-limited temporal resolution\cite{gierz2013snapshots,johannsen2013direct}. The specific sample used here was a quasi-freestanding monolayer graphene on 6H-SiC (0001)\cite{forti2011large}, showing a hole-pocket around the $\bar{K}$ point, as seen in Fig.\ref{fig:TR_GR}b). Similar to the case of WSe$_{2}$, graphene has a small real-space unit cell, which corresponds to a large first Brillouin zone in the reciprocal space, making it challenging to reach the zone boundary for most of the lab-based, non-HHG laser ARPES setups. Figure \ref{fig:TR_GR}a) illustrates the electronic response to the optical excitation, for which we utilized the idler mode of the OPA at 1.2 $\mu$m wavelength. Figure \ref{fig:TR_GR}a) is based on the momentum-integrated excitation spectrum of the graphene sample, plotted as a function of delay time between the pump- and probe-beam. The probe photon energy was 25.3~eV. The data was measured at room temperature and the acquisition time for each delay point was 5 minutes. In Fig.\ref{fig:TR_GR}b) we plot the electronic structure at the $\bar{K}$ point at different stages of time delays. The leftmost window shows the static spectrum whereas the middle and rightmost windows show the difference between the excited and static spectrum at $t_0$ and $t_0$+1~ps, respectively. The incidence angle of the probe beam is 15$^\circ$ upward in the vertical direction, and polarization is close to horizontal,  resulting in unequal intensity of the two branches\cite{mucha2008characterization}. The energy window indicated by the purple line on the right side of the Fig.\ref{fig:TR_GR}a) was integrated in energy and plotted in Fig.\ref{fig:TR_GR}c) as a function of delay time. The fit to the data (red line) is based on a 2-component exponential decay curve convoluted with a Gaussian. The $\tau_1$ and $\tau_2$ of the exponential decay function represent contributions from electron scattering with optical and acoustic phonons, respectively\cite{gierz2014non}. The overall temporal resolution is determined from the FWHM of the Gaussian distribution, which represents the temporal resolution broadening, while the $\tau$-parameters describe the rate of decay after the excitation. For the 7th harmonic (\textit{h$\nu$} = 25.3~eV), the system temporal resolution is determined to be $\sim$165 fs. Corresponding data recorded using 18.1~eV photons (not shown here) indicate a time resolution of $\sim$204~fs. In order to compare the time-bandwith product to that of a Fourier transform limited pulse, we combine these results with the energy resolution determined above and plot the energy resolution, $\Delta E$, as a function of the time resolution, $\Delta t$, in Fig.\ref{fig:TR_GR}d). Note that in the plot, the temporal contribution from the pump beam has been removed. Overall, the results show a $\Delta E \cdot\Delta t \sim 2400$~meV$\cdot$fs as compared to the transform limit of 1825~meV$\cdot$fs for a Gaussian pulse. This demonstrate that the present source is only 31\% above the transform limit. Several factors can contribute to the extra broadening both in time and in energy. Measured energy broadening can originate from space charge, analyzer resolution, stray fields and HHG conditions. Extra time broadening can, on the other hand, originate from the THG and HHG processes themselves, as well as from chirp induced by optical components, filters and windows. However, as noted previously, we do not observe any improved energy resolution with decreasing photon flux. Given the simulated analyzer energy resolutions for the settings used in our Fermi edge measurements (0.9 meV, 0.9 meV and 6.8 meV for photon energies 10.8 eV, 18.1 eV and 25.3 eV, respectively), we do not expect considerable analyzer contributions to the overall energy resolution and the determined time-bandwidth product is likely intrinsic to the light. The deviation from the ideal time-bandwidth product is not surprising as neither the time-bandwidth product of THG pulses nor the HHG generation conditions are expected to be ideal. Further improvements to both the drive pulses and the generation conditions could thus potentially improve the overall time-bandwidth product of the system further.

\section{\label{sec:Conclusion}Conclusion\protect }
In conclusion, we have designed a narrow bandwidth, high repetition rate XUV source for time-resolved ARPES. The available photon energies cover a wide range from 10.8~eV to 32.5~eV with an energy resolution of 9~meV at the lowest photon energy. The technical performance and suitability of the light source for time-resolved ARPES is demonstrated across test samples and typical quantum material systems such as gold, graphene, transition metal dichalcogenides and high temperature superconductors. The pump-line equipped with an OPA provides wavelength from 0.65~$\mu$m to 9~$\mu$m with pulse duration < 100~fs, allowing for both above-the-gap pumping and sub-gap pumping of coherent phonons across a wide range of materials. The combination of high repetition rate, wide range of photon energies and a continuously tunable wide-range of pump energies with a  time-of-flight detector, makes it possible to study the ultrafast dynamics over the whole first Brillouin zone in most crystalline materials.

\begin{acknowledgments}
This work was financially supported by the Knut and Alice Wallenberg foundation (No.2018-0104) and the Swedish research council VR (No.2019-00701). Q.G acknowledges the fellowship from China scholarship council (No.201907930007). M.D. acknowledges financial support from the Göran Gustafsson foundation. We thank Dr. Qiang Gao and professor Xingjiang Zhou from the institute of physics at Chinese Academy of Sciences, for providing the high-quality Bi-2212 sample, and professor Youguo Shi (IOP, CAS), for the high-quality WSe$_2$ sample.
\end{acknowledgments}

\section*{DATA AVAILABILITY}
The data that support the findings of this study are available
from the corresponding author upon reasonable request.

\nocite{*}
\bibliography{aipsamp}
\end{document}